\documentclass[3p]{elsarticle}

\pdfoutput=1

\usepackage{url}
\usepackage{mathtools}
\usepackage{xcolor}
\usepackage{booktabs}
\usepackage{amssymb}
\usepackage{algorithm}
\usepackage{subcaption}
\usepackage{algorithmicx}
\usepackage{algpseudocode}
\usepackage[hang,flushmargin]{footmisc}
\usepackage[normalem]{ulem}
\usepackage{hyperref}

\algnewcommand\algorithmicforeach{\textbf{for each}}
\algdef{S}[FOR]{ForEach}[1]{\algorithmicforeach\ #1\ \algorithmicdo}

\algnewcommand\algorithmicinput{\textbf{Input:}}
\algnewcommand\INPUT{\item[\algorithmicinput]}

\algnewcommand\algorithmicoutput{\textbf{Output:}}
\algnewcommand\OUTPUT{\item[\algorithmicoutput]}

\let\oldReturn\Return
\renewcommand{\Return}{\State\oldReturn}

\DeclarePairedDelimiter{\norm}{\lVert}{\rVert}

\DeclareMathOperator*{\argmin}{argmin}

\makeatletter
\def\ps@pprintTitle{%
	\let\@oddhead\@empty
	\let\@evenhead\@empty
	\def\@oddfoot{}
	\def\@oddhead{\bfseries\color{red}\uline{Published in ``Astronomy and Computing'' journal - DOI: 10.1016/j.ascom.2020.100364}}
	\let\@evenfoot\@oddfoot}
\makeatother

\newcommand{\z}[1]{\textcolor{black}{#1}}

\makeatletter
\newcommand{\algrule}[1][.2pt]{\par\vskip.5\baselineskip\hrule height #1\par\vskip.5\baselineskip}
\makeatother

\begin{document}
\begin{frontmatter}
\title{Optimal target assignment for massive spectroscopic surveys\tnoteref{mytitlenote}}
\tnotetext[mytitlenote]{This work was financially supported by the Swiss National Science Foundation (SNF) grant number 20FL21\_185771 and the SLOAN ARC/EPFL agreement number SSP523.}

\author{Matin Macktoobian$^{1}$}
\author{Denis Gillet$^{1}$}
\author{Jean-Paul Kneib$^{2}$}
\address{$^{1}$School of Engineering, Swiss Federal Institute of Technology in Lausanne (EPFL), 1015 Lausanne, Switzerland\\
	$^{2}$School of Basic Sciences, Swiss Federal Institute of Technology in Lausanne (EPFL), 1015 Lausanne, Switzerland}
\begin{abstract}
Robotics have recently contributed to cosmological spectroscopy to automatically obtain the map of the observable universe using robotic fiber positioners. For this purpose, an assignment algorithm is required to assign each robotic fiber positioner to a target associated with a particular observation. The assignment process directly impacts on the coordination of robotic fiber positioners to reach their assigned targets. In this paper, we establish an optimal target assignment scheme which simultaneously provides the fastest coordination accompanied with the minimum of colliding scenarios between robotic fiber positioners. In particular, we propose a cost function by whose minimization both of the cited requirements are taken into account in the course of a target assignment process. The applied simulations manifest the improvement of convergence rates using our optimal approach. \z{We} show that our algorithm scales the solution in quadratic time in the case of full observations. Additionally, the convergence time and the percentage of the colliding scenarios are also decreased in both supervisory and hybrid coordination strategies.
\end{abstract}


\end{frontmatter}

\section{Introduction}
\interfootnotelinepenalty=10000 
Modern observational astronomy is interested in the study of the dark matter distribution in the universe to investigate its evolution. For this purpose, cosmological spectroscopy has been taken into account to generate the map of the observable universe. In particular, electromagnetic waves emanating from target galaxies are collected using fibers. Then, the received signals are processed using a spectrograph to compute a desired map. The observable universe is about to be divided into specific sets of observations each of which includes a multitude of galaxies. Then, the data obtained from various surveys will generate a unified map depicting the observable universe.  

Many spectroscopic surveys are defined to be pursued based on a class of projects such as SDSS-V\cite{kollmeier2017sdss}. These projects seek to develop the ground telescopes equipped with fibers and spectrographs such as DESI \cite{flaugher2014dark}, MOONS \cite{cirasuolo2014moons}, PFS \cite{takada2014extragalactic}, SLOAN \cite{york2000sloan}, etc. In these telescopes, a robotic fiber positioner \cite{horler2018robotic} is attached to each optical fiber. Thus, a fiber can rotationally move in the workspace of its associated robotic fiber positioner to reach various spots associated with observation targets. The overall collection of robotic fiber positioners is located at a particular area of the telescope called focal plane. 

The spectroscopic survey generation is typically done in $3$ phases. Namely taking a specific observation into account, one first has to assign each target of an observation to a specific robotic fiber positioner of a planned telescope (\textit{assignment phase}). Then, the robotic fiber positioners have to be coordinated to reach their assigned targets such that no collision occurs during the \textit{coordination phase}\cite{macktoobian2019sdss,macktoobian2019complete}. Finally, the collected spectroscopic signals are sent to a spectrograph to generate the map of the observation (\textit{processing phase}). Because of the complexity of the system and the dense placement of robotic fiber positioners, the coordination phase is often challenging to be handled. In particular, one is interested in fast collision-free coordinations whose convergence rates\footnote{Given a set of robotic fiber positioners, convergence rate is defined as the number of those which reach their desired position at the end of their coordination divided by their overall number.} are maximized. One may note that the assignment phase dramatically influences the achievement of the quoted favorite coordination. Thus, any improvement of the assignment process may directly contribute to the realization of safer collision-free and faster coordinations.

As already stated, each observation often requires a large number of objects to be assigned to a set of robotic fiber positioners \cite{macktoobian2019heterogeneous}. So, a set of methodologies were proposed to handle assignments. For example, random assignment simply assigns each target to a random unassigned robotic fiber positioner. This method takes no specific criteria into account, thereby being computationally plausible. However, the coordination phase may become very challenging because random assignment may assign some targets to a dense neighborhood of robotic fiber positioners. In this case, the coordination phase may struggle to simultaneously reach proper convergence rates and avoid potential collisions between the robotic fiber positioners. \z{Flow-based assignment \cite{blanton2003efficient} was another scheme which solves the target assignment problem as a network flow one. The method aims to maximize the number of targets to which fibers are assigned. This strategy uses network flow graphs to solve the problem. Namely, it first identifies a set of fibers which do not collide any other peers. Then, the set of unallocated fibers are taken into account to resolve collisions using a network flow graph. This method was basically designed to address target assignment in non-robotic focal planes. In other words, however collisions among fibers are addressed by the method, it may not be able to properly handle collisions among robotic fiber positioners. In automatized focal planes, collisions depend on real-time states of robotic fiber positioners during coordinations. So in practice, it is unlikely to be able to consider all potential collisions corresponding to numerous coordination solutions of a system of robotic fiber positioners in the assignment phase. Thus, modular perspective in decoupling assignment phase from coordination phase may be more effective to avoid collisions and to find fast and safe coordinations during coordination phases.} Later, the drainage algorithm \cite{morales2012fibre} was proposed based on a tiling approach. This method ensures that the maximum number of targets are observed in an observation. In particular, targets are moved among various lists of unassigned robotic fiber positioners each of which can observe a specific set of targets. Finally, each target is assigned to the shortest list including observable targets. This method assimilates the target-to-positioner ratio corresponding to the overall desired targets with respect to robotic fiber positioners. As a disadvantage, this strategy assumes no physical size for robotic positioners, thereby neglecting any potential collisions between them.

Alternatively, target-based assignment method \cite{schaefer2016target} assigns positioners to targets instead of targets to positioners. The assignment ratio of this algorithm is improved compared to that of the drainage algorithm. However, this method is  computationally more intensive than the drainage algorithm. The target-based assignment method seeks the ease of path finding, say, decreasing the occurrences of collisions\footnote{A collision occurs when two neighboring robotic fiber positioners violently strike against another.} and deadlocks\footnote{A deadlock is a situation in which two neighboring robotic fiber positioners, which blocks each other's paths through their target spots, stop moving at a location which is not their target spots. Each back-and-forth movement of such a pair of robotic fiber positioners is counted as one deadlock.} using a parity-based mechanism. In particular, each robotic fiber positioner can rotate in two different directions one of which may be less susceptible to collisions and deadlocks. Thus, one may set and vary parities associated with a robotic fiber positioners set to find a parity set according to which the robotic fiber positioners can converge to a specific configuration with less colliding and deadlock issues. This method checks the existence of conflicts at the assignment time. Thus, achieving the least collision-prone solution is not guaranteed. Nonetheless, there is no guideline based on which one could set parities in a systematic way to achieve the reported $\sim2\%-3\%$ improvement rate in assignment gain. Thus, none of the methods above simultaneously minimizes both the required coordination and the occurrence of collisions and/or deadlocks. 

In this paper, we find an optimal solution to the target assignment problem which minimizes the required coordination and maximizes the distance between assigned positioners to minimize collisions and/or deadlocks. In particular, we embed the quoted criteria into a cost function. Then, we propose an algorithm which finds the optimal target-to-positioner assignments constrained to the cost function minimization. We show that our quadratic algorithm is computationally efficient enough to solve the target assignment problem associated with crowded robotic fiber positioners sets.

Our method improves the performances of both hybrid \cite{tao2018priority} and supervisory \cite{macktoobian2019supervisory} coordination approaches. In particular, hybrid coordination reaches higher convergence rates using our optimal assignment strategy, The supervisory coordination, which seeks complete coordination of a robotic fiber positioners set, often suffers from computational complexity issues. To be specific, supervisory coordination  requires intensive computations to solve the completeness problem if the random, drainage, or target-based algorithms are used in the assignment process. In contrast, we observe that optimal target assignment efficiently reduces the state space size of coordination supervisors. Put differently, optimal target assignment surpasses the methods quoted above to solve the coordination problem in a shorter time. 

The remainder of the paper is organized as follows. Optimal target assignment process is presented in Section \ref{sec:OTA}. In particular, a cost function is defined to address the requirements of the problem solution. Then, an algorithm is established to solve the problem. We also demonstrate the quadratic complexity of our algorithm. Section \ref{sec:sim} illustrates how our optimal assignment algorithm improves the coordination results of both hybrid and supervisory methods. Section \ref{sec:conc} summaries our accomplishments.
\begin{figure*}[t!]
	\centering
	\hspace*{-2mm}\begin{subfigure}[t]{0.5\textwidth}
		\includegraphics[scale=0.6]{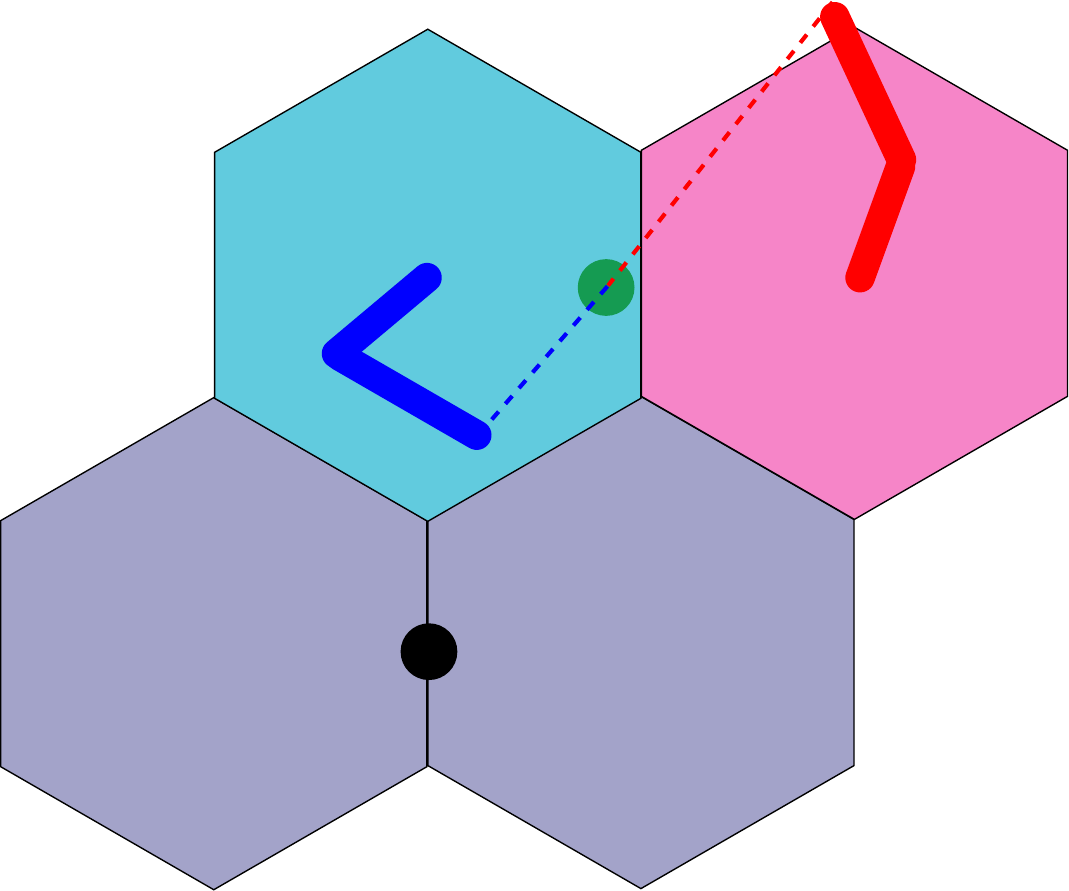}
		\caption{\z{The blue robotic fiber positioner fulfills the minimum coordination criterion.}}
		\label{fig:min}
	\end{subfigure}%
	\hspace*{2mm}\begin{subfigure}[t]{0.5\textwidth}
		\includegraphics[scale=0.6]{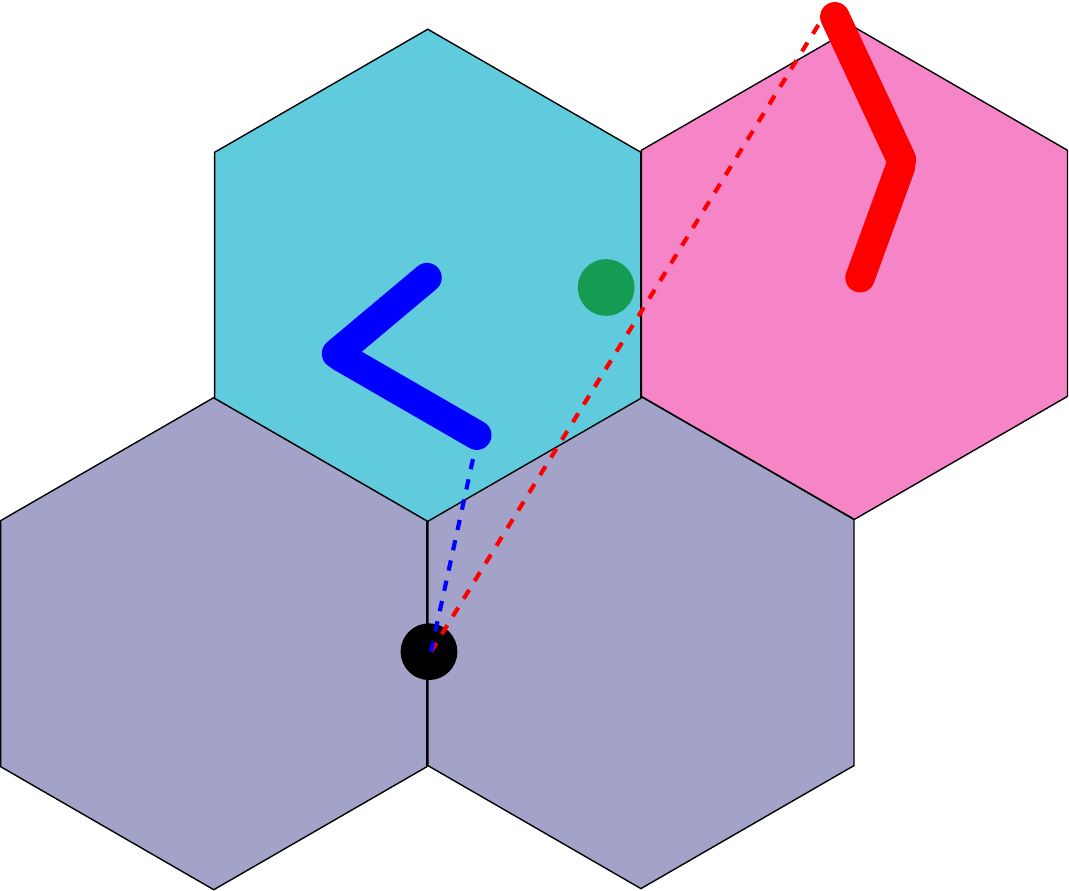}
		\caption{\z{The red robotic fiber positioner satisfies the maximum distribution criterion.}}
		\label{fig:max}
	\end{subfigure}
	\caption{\z{Optimization requirements evaluations in a typical assignment scenario. (As a patch of a typical focal plane, the gray hexagons represent robotic fiber positioners which have been already allocated, and the black circle depicts their average coordinate. Two unallocated robotic fiber positioners, i.e., the dark blue and red manipulators, are confined to their corresponding hexagons. Each of the robotic fiber positioners can reach the green circle representing an unassigned target.)}}
\end{figure*}
\section{Optimal Target Assignment}
\label{sec:OTA}
In this section, we start by defining a cost function which takes the trade-off between the following requirements into account.
\begin{enumerate}
	\item the minimum coordination of the robotic fiber positioners which reach a particular set of targets: each target should be assigned to the robotic fiber positioner whose initial ferrule's coordinate is in the closest distance to the target's projected location on the host focal plane compared to the projected locations of the reminder of the reaching robotic fiber positioners. \z{(see, Fig. \ref{fig:min})}  
	\item the maximum distribution of the robotic fiber positioners which reach a particular set of targets: each target should be assigned to the robotic fiber positioner which is located at the farthest distance from the already-allocated robotic fiber positioners. \z{(see, Fig. \ref{fig:max})}
\end{enumerate}
We first define the \textit{reachability relation} which specifies a robotic fiber positioner reaching a particular target. In particular, let $\pi$ and $t$ be a robotic fiber positioner and a target, respectively. Then, the reachability relation $\mathcal{R}(\pi,t)$ holds if $t$ is reachable by $\pi$. We also specify the set of all robotic fiber positioners reaching a specific target as its \textit{reachable set}. Namely, let $\mathcal{P}$ and $t$ be a set of robotic fiber positioners and a target, respectively. Then, $\Gamma_{t}^{\mathcal{P}}$ is the largest subset of $\mathcal{P}$ whose elements reach $t$ as follows.
	\begin{equation}
	\begin{split}
	\Gamma_{t}^{\mathcal{P}} := & \{\pi \in \mathcal{P}|\mathcal{R}(\pi,t) \}\\
	:= & \{{}^{i}\Gamma_{t}^{\mathcal{P}}|1 \le i \le |\mathcal{P}| \} 
	\end{split}
	\end{equation}
We are interested in a measure to represent the distance of a target from a set of robotic fiber positioners. Thus, we define the \textit{average coordinate} associated with a batch of robotic fiber positioners. Given an index set $\mathcal{I}$ and a set of robotic fiber positioners $\mathcal{P} := \{\pi_{i}|i \in \mathcal{I} \}$, let $\mathcal{Q} := \{q_{i}|i \in \mathcal{I} \}$ be the coordinate set corresponding to $\mathcal{P}$. Then, the average coordinate $\overline{\mathcal{Q}}$ associated with $P$ is defined as below.
	\begin{equation}
	\overline{\mathcal{Q}} := \frac{\sum\limits_{i \in \mathcal{I}} q_{i}}{|\mathcal{I}|}
	\end{equation}
	Given a target $t$, $q_t$ represents the \textit{projected coordinate} of $t$ on the focal plane corresponding to the set of robotic fiber positioners which can observe $t$.
Now we are set to mathematically formulate our requirements. In particular, we take the minimum coordination criterion into account corresponding to a target $t$ by assigning the positioner $\pi_{i} \in {}^{i}\Gamma^{\mathcal{P}}_{t}$ to it which minimizes the following subfunction\footnote{Operator $\norm{\cdot}_2$ represents Euclidean norm.}
\begin{equation}
\left\Vert q_{i}-q_{t}\right\Vert_{2}.
\end{equation}
Furthermore, we consider the maximum distribution criterion for a target $t$ by assigning the positioner $\pi_{i} \in {}^{i}\Gamma^{\mathcal{P}}_{t}$ to it which maximizes the following subfunction
\begin{equation}
\label{eq:MD}
\sum\limits_{j \in \mathcal{J}}\left\Vert q_{i}-q_{j}\right\Vert_{2}.
\end{equation}
Here $J$ is an index set \z{corresponding to the set of already allocated robotic fiber positioners}. 

One may note that the maximum distribution criterion is computationally labor-intensive. So, we present a lower bound for this criterion which is computationally more plausible. In particular, let $\mathcal{P}$ and $t$ be a set of robotic fiber positioners and a target to be observed, respectively. Denote by $\overline{\mathcal{Q}}$ the average coordinate associated with $\mathcal{P}$. According to the extended triangle inequality, we have
\begin{equation}
\sum\limits_{j \in \mathcal{J}}\left\Vert q_{i}-q_{j}\right\Vert_{2} \ge \left\Vert\sum\limits_{j \in \mathcal{J}}\bigl(q_{i}-q_{j}\bigr)\right\Vert_{2},
\end{equation}
which gives
\begin{equation}
\begin{split}
\left\Vert\sum\limits_{j \in \mathcal{J}}\bigl(q_{i}-q_{j}\bigr)\right\Vert_{2} = &  \left\Vert\vert\mathcal{J}\vert q_{i}-\sum\limits_{j \in \mathcal{J}}q_{j}\right\Vert_{2}\\
=&\vert\mathcal{J}\vert\left\Vert q_{i}-\frac{\sum\limits_{j \in \mathcal{J}}q_{j}}{\vert\mathcal{J}\vert}\right\Vert_{2}\\
\ge & \left\Vert q_{i}-\overline{\mathcal{Q}}\right\Vert_{2}.
\end{split}
\end{equation}
Therefore, a lower bound for the maximum distribution criterion, i.e., (\ref{eq:MD}), is
	\begin{equation}
	\left\Vert q_{i}-\overline{\mathcal{Q}}\right\Vert_{2}.
	\end{equation} 
Thus, the overall cost function, the minimization of which simultaneously satisfies the both optimal criteria, is as follows.
\begin{equation}
\label{eq:opt}
\frac{\left\Vert q_{i}-q_{t}\right\Vert_{2}}{\left\Vert q_{i}-\overline{\mathcal{Q}}\right\Vert_{2}} \quad (\forall i : 1 \le i \le \norm{\Gamma^{\mathcal{P}}_{t}})
\end{equation} 
The optimal assignment solver (OAS) algorithm takes (\ref{eq:opt}) into account to solve the target assignment optimization problem. 
\begin{algorithm}[H]
	\renewcommand{\thealgorithm}{}
	\caption{Optimal Assignment Solver (OAS)}
	\begin{algorithmic}[1]
		\INPUT 
		\Statex $\mathcal{P}$ \Comment Robotic fiber positioners set
		\Statex $\mathcal{T}$ \Comment Targets set
		\OUTPUT 
		\Statex Assigned positioner-target tuples
		\algrule[1pt]
		\State $\mathcal{Q} \leftarrow \varnothing$
		\State $\text{sort } \mathcal{T} \text{ based on the observation priorities in descending order}$
		\ForEach {$t \in \mathcal{T}$}
		\State assign $t$ to $\pi_{i}$ such that $ i = \argmin\limits_{i}\frac{\left\Vert q_{i}-q_{t}\right\Vert_{2}}{\left\Vert q_{i}-\overline{\mathcal{Q}}\right\Vert_{2}} \newline\hspace*{2mm} \quad (\forall i : 1 \le i \le \norm{\Gamma^{\mathcal{P}}_{t}})$
		\State $\mathcal{P} \leftarrow \mathcal{P}\setminus\{\pi_{i}\}$
		\State $\mathcal{Q} \leftarrow \mathcal{Q}\dot{\cup}\{\pi_{i}\}$\Comment{Symbol $\dot{\cup}$ denotes disjoint union operation.}
		\EndFor
		\Return The assigned positioner-target tuples
	\end{algorithmic}
\end{algorithm}
Now, we demonstrate the quadratic computational complexity of the OAS algorithm. In particular, let $\mathcal{T}$ and $\mathcal{P}$ be a set of targets and robotic fiber positioners, respectively. Given $n:= |\mathcal{T}|$ and $k:= |\mathcal{P}|$, the sorting process can be executed in $\mathcal{O}(n\log n)$, which is not the computational bottleneck of the algorithm. Then, the loop has to check a subset of robotic fiber positioners for each target. In the worst case, the $i$\textsuperscript{th} iteration of the loop has to check $k-i+1$ robotic fiber positioners with respect the $i$\textsuperscript{th} target. Thus, the overall number of loop operations is
\begin{equation}
nk - (1+2+\cdots+n)+2 = nk - \dfrac{n(n+1)}{2} +2.
\end{equation}
In the worst case, we ideally suppose $k=n$ to maximize the information obtained from an observation. It turns out that the complexity is $\mathcal{O}(n^2).$ Thus, overall computational complexity of the OAS algorithm is quadratic.
\section{Simulations}
\label{sec:sim}
In this section\footnote{We conduct the tests on a ASUS ZenBook UX410UAR with an Intel Core i7-8550U @ 1.8GHz x 4 processor, Intel UHD Graphics 620 graphic card on an Microsoft Windows 10, 10.0.15063 version.}, we observe how our optimal target assignment method improves the performance of the overall coordination of robotic fiber positioners set in view of different measures. In particular, we compare what our strategy achieves to those of the target-based assignment, known as \z{one of the most promising} available assignment methods up to now. \z{The target distributions are assumed to be random. We performed $50$ simulated scenarios per each population of robotic fiber positioners whose overall averages are presented below.}  

Major parameters of the simulation are specified in the Table \ref{tab:tbl}.
\begin{table}[H]
	\begin{center}
		\caption{The parameters applied to the simulations}
		\label{tab:tbl}
		\begin{tabular}{ll}
			\toprule
			\textbf{Parameter} & \textbf{Value}\\
			\midrule 
			The length of the first arm & $8.000$ mm\\
			The length of the second arm & $17.000$ mm\\The width of the first arm & $8.000$ mm\\
			The width of the second arm& $4.000$ mm\\
			The ferrule size & $1$ mm\\
			The maximum speed of the first arm's actuator & $30$ rpm\\
			The maximum speed of the second arm's actuator & $20$ rpm\\
			The time step of control command generation & $10$ ms\\
			\bottomrule 
		\end{tabular}
	\end{center}
\end{table}

\begin{figure}
	\centering
	\includegraphics[scale=0.7]{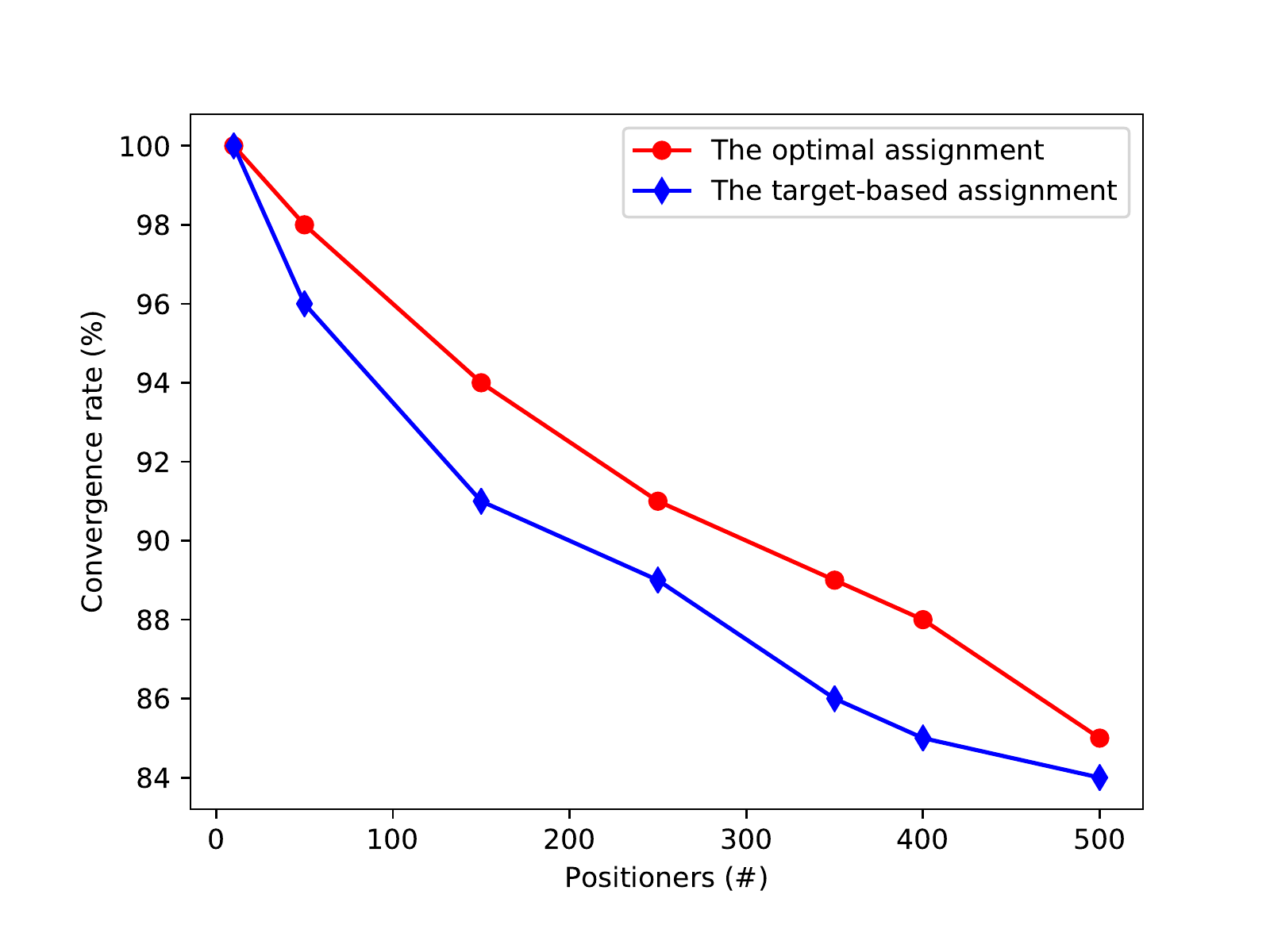}
	\caption{The convergence rate comparison between optimal and target-based assignments during hybrid coordinations}
	\label{fig:rate}
\end{figure}

The hybrid coordination \cite{tao2018priority} generally does not guarantee the complete coordination of a typical robotic fiber positioners set. Thus, the convergence rate is a measure to assess the effectiveness of a coordination process. In particular, Fig. \ref{fig:rate} illustrates the convergence rates corresponding to various numbers of robotic fiber positioners. One observes that the hybrid coordination method yields higher convergence rates using our optimal target assignment method.

\begin{figure}[H]
	\centering
	\includegraphics[scale=0.7]{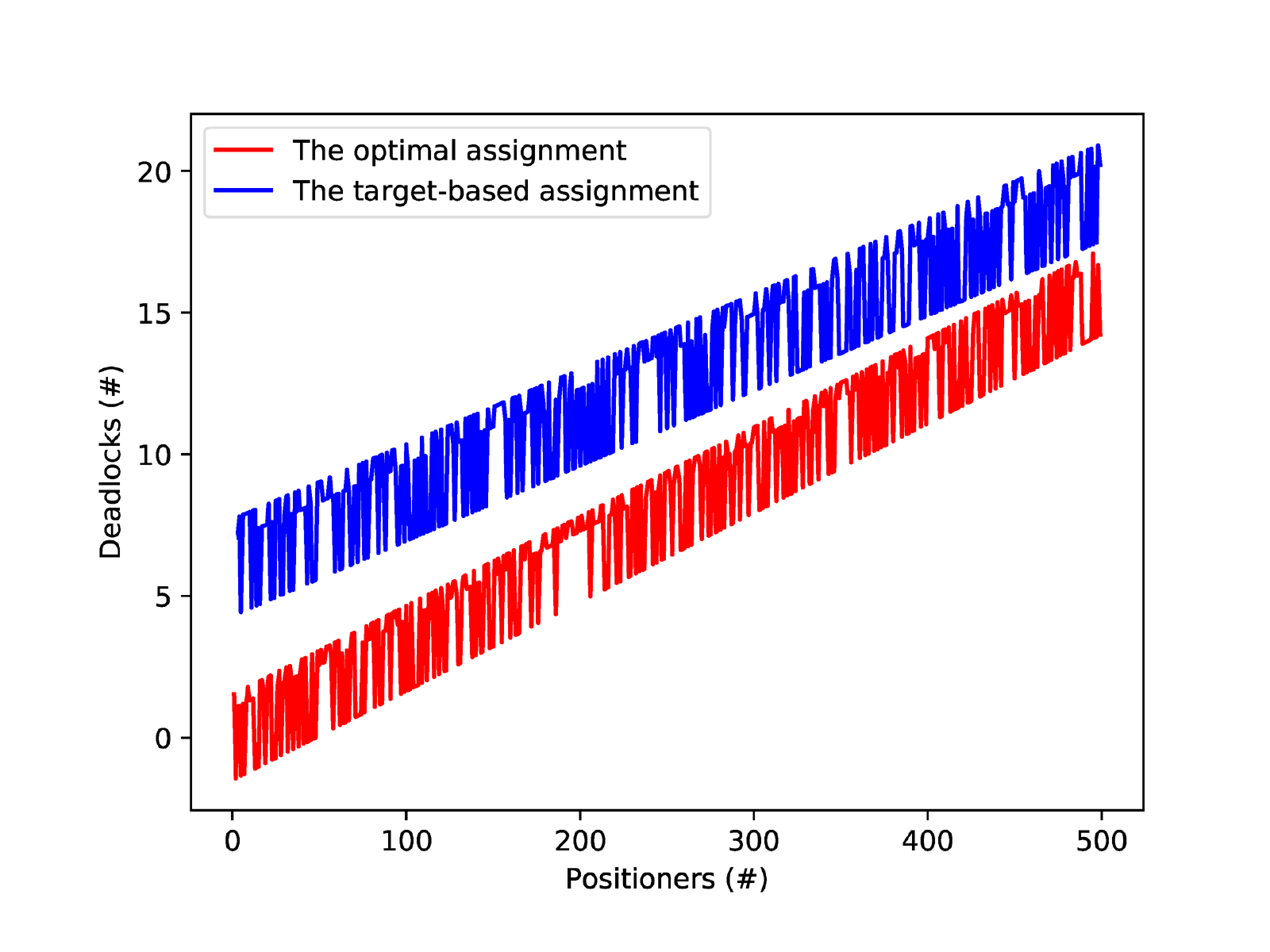}
	\caption{The deadlock occurrences using optimal assignment during hybrid coordinations}
	\label{fig:deadlocks}
\end{figure}

The number of deadlocks during a coordination process is correlated to the required time to reach the coordinated configuration. Put differently, the less deadlocks occur in the course of coordination, the faster the final coordinated configuration is achieved. In particular, Fig. \ref{fig:deadlocks} depicts the decrement of deadlocks during coordinations when optimal target assignment is taken into account.

\begin{figure}[H]
	\centering
	\includegraphics[scale=0.7]{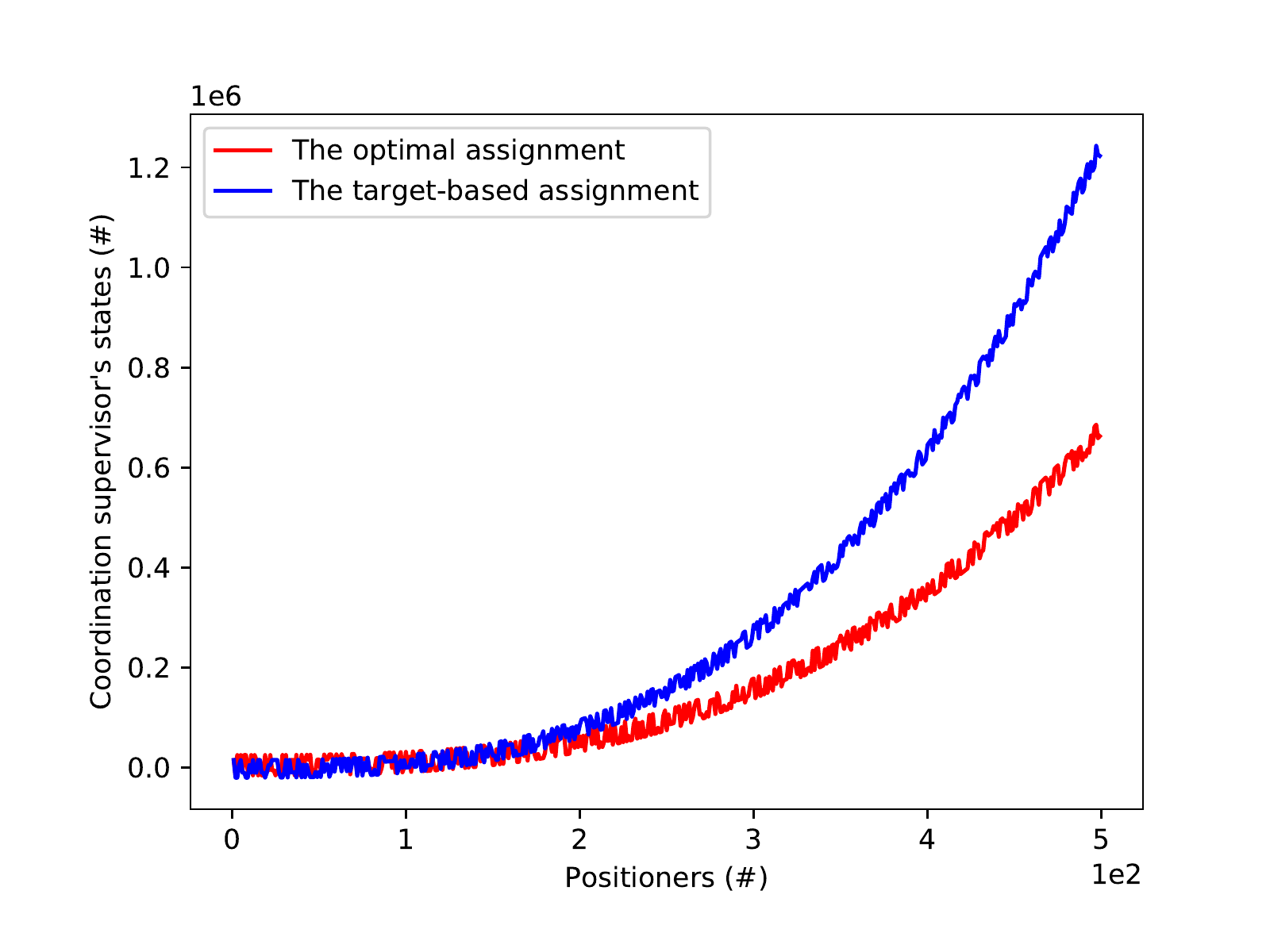}
	\caption{The decrement of coordination supervisors' state space size during supervisory coordinations}
	\label{fig:states}
\end{figure}

The supervisory coordination \cite{macktoobian2019supervisory} is also improved using the optimal target assignment. Namely, a coordination supervisor is a finite state machine some of whose strings are solutions to the complete coordination problem associated with a specific set of robotic fiber positioners. Since the solutions have to be found through the overall structure of a coordination supervisor, the state size of the coordination supervisor significantly impacts on the time required to solve a complete coordination problem. In particular according to the Fig. \ref{fig:states}, the optimal target assignment method resembles the target-based one in view of the state space size of coordination supervisors if the overall robotic fiber positioners sets are not very populated. However, we observe that the optimal target assignment method surpasses the target-based counterpart regarding this factor when the number of robotic fiber positioners is increased. 
\section{Conclusion}
\label{sec:conc}
This paper introduces a new method to improve the state-of-the-art target assignment algorithm based on an optimization perspective. In particular, the optimal target assignment takes a cost function into account according to which the distribution density of assigned robotic fiber positioners is locally decreased, thereby less frequent occurrences of deadlocks during coordinations. We also minimize the cost function seeking the target-fiber pairs whose Euclidean distance is as small as possible. Thus, the overall coordination time is minimized. We also show that our optimization algorithm is efficiently scalable in quadratic time. The applied simulations represent the efficiency of the optimal target assignment scheme to improve the performances of both the hybrid and the supervisory coordination techniques.

\bibliographystyle{elsarticle-num}
\bibliography{references}

\end{document}